# A Quantum Scattering Interferometer

Russell A. Hart, Xinye Xu,[*] Ronald Legere,[†] & Kurt Gibble
*Department of Physics, The Pennsylvania State University, University Park, PA 16802, USA.*

**The collision of two ultra-cold atoms results in a quantum-mechanical superposition of two outcomes: each atom continues without scattering and each atom scatters as a spherically outgoing wave with an s-wave phase shift. The magnitude of the s-wave phase shift depends very sensitively on the interaction between the atoms. Quantum scattering and the underlying phase shifts are vitally important in many areas of contemporary atomic physics, including Bose-Einstein condensates,[1-5] degenerate Fermi gases,[6-9] frequency shifts in atomic clocks,[10-12] and magnetically-tuned Feshbach resonances.[13] Precise measurements of quantum scattering phase shifts have not been possible until now because, in scattering experiments, the number of scattered atoms depends on the s-wave phase shifts as well as the atomic density, which cannot be measured precisely. Here we demonstrate a fundamentally new type of scattering experiment that interferometrically detects the quantum scattering phase shifts of individual atoms. By performing an atomic clock measurement using only the scattered part of each atom, we directly and precisely measure the difference of the s-wave phase shifts for the two clock states in a density independent manner. Our method will give the most direct and precise measurements of ultracold atom-atom interactions and will place stringent limits on the time variations of fundamental constants.[14]**

In our experiment, we juggle atoms[15] in an atomic fountain clock by upwardly tossing two gaseous clouds of caesium atoms in rapid succession by laser-cooling them in a frame that moves upwards at 2.5 to 3.4 m/s. Gravity slows the atoms and, after the first cloud reaches its apogee, the two clouds pass through one another and the atoms collide. For a short time delay between launches, of order $\Delta t$=10 ms, the relative velocity of the atoms is about $v_r$=g $\Delta t$ =10 cm/s, corresponding to a collision energy of $E = mv_r^2/4 = 40\mu K \times k_B$, which is much greater than their temperature of 250 to 510 nK. We prepare the atoms in Cloud 1 in a pure $|F,m\rangle$ state (e.g. $|4,4\rangle$) and Cloud 2 in one of the clock states ($|3,0\rangle$). Both clouds pass through a microwave (clock) cavity which puts the atoms in Cloud 2 in a coherent superposition of the two clock states, $|3,0\rangle$ and $|4,0\rangle$. The phase of this coherence precesses at the caesium clock frequency, $\nu \simeq 9.2$ GHz.[12] When the two clouds collide, the atoms scatter as illustrated in Fig. 1. The s-wave part of the atomic wavefunction in each clock state, $|3,0\rangle$ or $|4,0\rangle$, scatters off of the atoms in Cloud 1, acquiring an s-wave phase shift, $\delta_3$ or $\delta_4$. After scattering, the atoms fall back through the microwave cavity, which converts the phase difference between the clock coherence and the microwave field into a population difference of the clock states, which we detect. This population difference for the unscattered part of each atom yields the usual transition probability for a clock as a function of microwave frequency, known as Ramsey fringes[12] (diamonds in Fig. 2a). Here we instead detect only a scattered part of each atom's wavefunction, for which the phase of the coherence is shifted by the difference of the s-wave phase shifts, $\Phi=\delta_3 - \delta_4$ (circles in Fig. 2a). In this way, we use atomic-clock interferometry to directly observe the difference of the s-wave phase shifts. To demonstrate this technique, we scatter the caesium clock states off of $|4,4\rangle$ at $v_r$ =9.92 cm/s and measure $\Phi$=–0.141(8) radians.

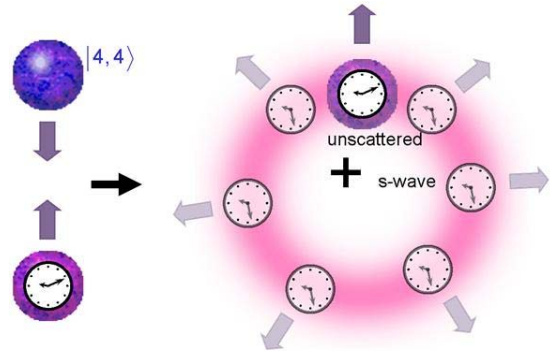

**Fig. 1. Schematic of the Experiment.** We collide an atom in a coherent superposition of the two caesium clock states with a caesium atom in a pure $|F,m\rangle$ state, such as $|4,4\rangle$. When the clock states scatter, they experience different s-wave phase shifts, shifting the phase of the clock coherence by the difference of the s-wave phase shifts. By detecting only the scattered part of each clock atom, we directly observe the difference of the s-wave phase shifts.

To select a clock atom that scatters, we use the Doppler shift and a narrow two-photon Raman transition.[15-17] In Fig. 3, we show velocity distributions of the vertical velocity component of Cloud 2, prepared in $|3,0\rangle$, when it collides with $|4,4\rangle$ atoms in Cloud 1. For the data in Fig. 3, the microwave pulses to the clock cavity are disabled so that Cloud 2 is not prepared in a coherent superposition of the clock states. The velocity selective probe pulse transfers atoms from $|3,0\rangle$ to $|4,0\rangle$ with a bandwidth of 1.4 cm/s and we detect the number of atoms in F=4. Before the probe pulse, we push the atoms in F=4 from the fountain with a laser beam tuned to excite F=4 atoms.[15,17] We push the F=4 atoms either early, before Cloud 1 enters the clock cavity, or late, right after both clouds return downwardly through the clock cavity. Early clearing gives the "no-collisions" signal in Fig. 3a (violet) and late clearing allows the two clouds to collide before we clear Cloud 1, giving the "collisions" signal in Fig. 3a (aqua). In the magnified Fig. 3b, the difference between the "collisions" and "no-collisions" curves between $v_z = -5$ cm/s and 2 cm/s represents scattered atoms (Fig. 3c). For both curves in Figs. 3a & b, we subtract the small backgrounds in Fig. 3d, obtained by inhibiting the preparation of Cloud 2 in $|3,0\rangle$. To observe the Ramsey fringes of scattered atoms (circles in Fig. 2), we fix the probe velocity at $v_z=0$ in Fig. 3c, which corresponds to 90° scattering, enable the microwave clock pulses, and then scan the frequency of the microwave clock pulses. The contributions to the Ramsey fringes at $v_z=0$ from the "no-collisions" and backgrounds are small, with phase shifts consistent with 0 (Fig. 2b & c). For this first demonstration, we choose 90° scattering to avoid contributions from p-waves.

---

[*] Present address: Department of Physics, East China Normal University, Shanghai 200062, China
[†] Present address: MIT Lincoln Laboratory, Lexington, MA 02420, USA

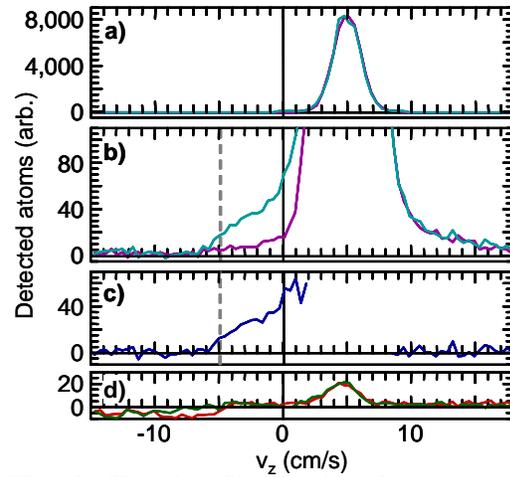

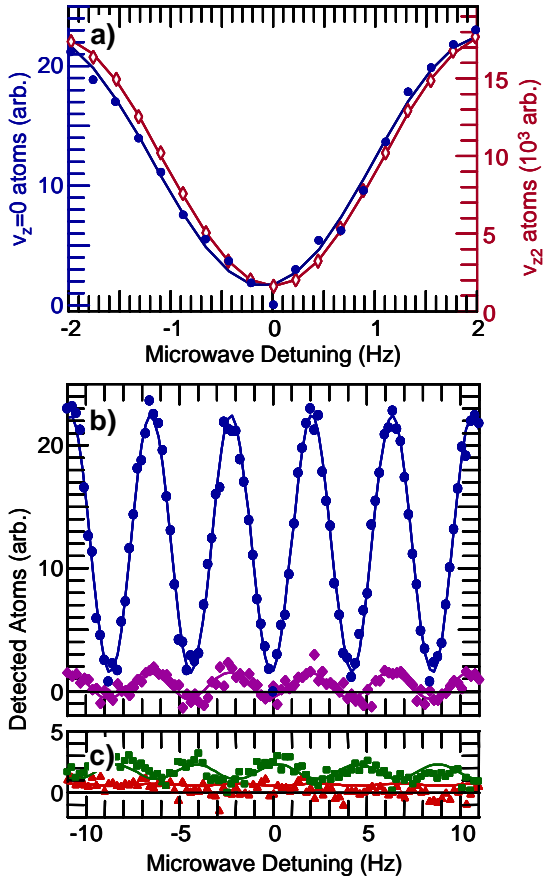

**Fig. 2. Ramsey fringes for scattered and unscattered atoms.** a) The central Ramsey fringe for clock atoms that have s-wave scattered from $|4,4\rangle$ atoms at 90° ($v_z=0$, circles) with $v_r=9.92$ cm/s. The reference Ramsey fringe is from the unscattered clock atoms in Cloud 2 ($v_{z2}$, diamonds). The fringes for the scattered atoms have a phase shift of −0.141 radians, which is the difference of the s-wave phase shifts. b) The entire Ramsey pattern for scattered atoms at $v_z=0$ (blue) and for "no-collisions" (diamonds) as in Fig. 3. c) The background Ramsey fringes as in Fig. 3. for "collisions" (squares) and "no-collisions" (triangles). For this data, the difference of the s-wave phase shifts for the clock states scattering off of $|4,4\rangle$ is relatively small, $\Phi=-0.141(12)$ radians. Each circle represents the average of four differences of four measurements, requiring 16 cycles of the atomic fountain. The central Ramsey fringe is a minimum because the atoms begin in the $|3,0\rangle$ clock state and we detect the final number in $|3,0\rangle$. In a) there are nearly 1,000 times more unscattered atoms (maroon) than detected scattered atoms (blue).

**Fig. 3. Velocity distribution for atoms in Cloud 2.** Cloud 2 is prepared in $|3,0\rangle$ and Cloud 1 in $|4,4\rangle$, with $v_r=9.92$ cm/s. In this centre-of-mass frame, the vertical velocity component of Cloud 1 is $v_{z1}=-4.96$ cm/s and $v_{z2}=4.96$ cm/s for Cloud 2. a) The aqua (violet) curves show the raw velocity distribution of Cloud 2 for "collisions" ("no-collisions") when we clear Cloud 1 from the fountain late (early). b) Magnification of a) by a factor of 100. The difference of the "collisions" and "no-collisions" curves, plotted in c), between $v_z=-5$cm/s and 2 cm/s represents scattered atoms. d) Background for "collisions" (green) and "no-collisions" (red). In Fig. 2, the blue (maroon) Ramsey fringes are for $v_z=0$ ($v_{z2}$). About 0.1% of the atoms scatter into the 1.4 cm/s detected velocity width at 90°.

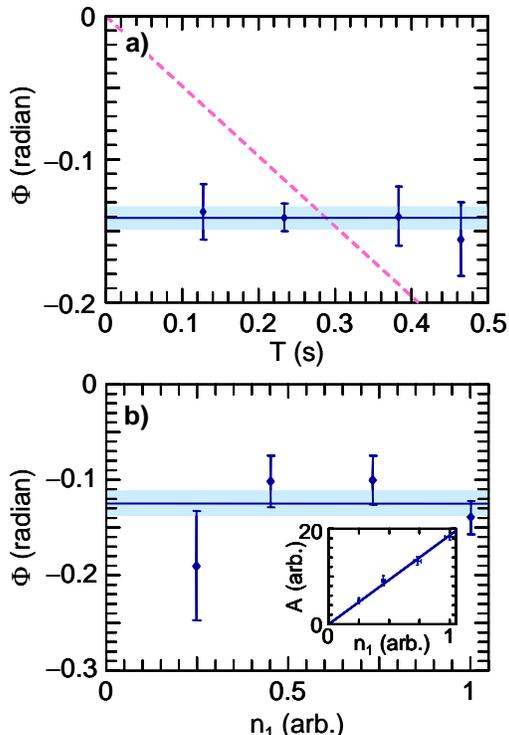

**Fig. 4. The s-wave phase shift of scattered atoms.** a) The phase shift versus the interrogation time T between the two microwave field interactions. The phase shift is independent of T, as opposed to being proportional to T as it is for a frequency shift (dashed). We increase T by increasing the launch velocity in the fountain. The best fit to all the data is $\Phi=-0.141(8)$. b) The phase shift and Ramsey fringe amplitude A (inset) versus the density of Cloud 1. The phase shift is independent of the density rather than proportional to density, as is the usual cold-collision frequency shift in clocks. As expected, the Ramsey fringe amplitude is proportional to the density of atoms in Cloud 1. The error bars and bands represent standard errors.

These measurements are qualitatively different than the usual frequency shift in an operating atomic clock due to cold collisions. The usual frequency shift arises from a collision rate which gives the rate of shifting the phase of the coherence, which is simply a frequency shift. The usual frequency shift is due to the quantum-mechanical interference in the forward direction between the unscattered and scattered parts of each atom[18,19] and is proportional to density.[10-12] Here, the Ramsey fringes for the atoms that scatter have a phase shift (which is independent of density) instead of a frequency shift because the coherence of the scattered part of each atom experiences both s-wave phase shifts, $\delta_3$ and $\delta_4$. To demonstrate the qualitative differences, we show in Fig. 4a the phase shift as a function of the free precession time T between the two microwave pulses for T=0.115 s to 0.450 s. We increase the free precession time by increasing the launch velocity of both clouds, so that their apogees above the clock cavity are higher.[12] The phase shift is independent of T. It is clearly inconsistent with a frequency shift, for which the phase shift would increase linearly with T (dashed line). Further, the magnitudes of the effects are dramatically different. The largest cold-collision frequency shift that has been observed in a clock is −5.5 mHz.[10] Here, for T=0.115 s, a phase shift of $\Phi=-0.141$ rad is, expressed as a frequency shift, −200 mHz.

A key feature of this technique is that this difference of quantum scattering phase shifts is independent of the atomic density to lowest order (Fig. 4b). Many experiments have probed a wide variety of scattering effects, including velocity redistribution,[17,20] frequency shifts,[10,11] and inelastic losses.[21] Each of these effects is given by a rate $n\, v_r\, \sigma$ where $n$ is the atomic density and $\sigma$ is the cross section for the process. Since the best measurements of the density of cold atoms do not achieve even 1% accuracy,[22,23] it is generally not possible to precisely determine atomic scattering phase shifts and the atom-atom interactions from these measurements. Indeed for caesium, many scattering results appeared thoroughly inconsistent until the spectroscopic observation of many Feshbach resonances,[24] combined with a theoretical analysis, determined the caesium interaction properties.[25] Here, although the number of scattered atoms and the amplitude of the Ramsey fringes are proportional to the density of Cloud 1 (inset Fig. 4b), the phase shift is independent of the atomic density and therefore can utilize the high accuracy available with atomic clock techniques. In this first measurement, our statistical uncertainty is 6% (of even a relatively small difference of s-wave phase shifts), compared to typical density uncertainties of a factor of two. While measurements of differential cross sections also yield density independent phase shift differences,[15,26,27] potential systematic errors that depend on the scattering angle inhibit the precision. Future improvements of our precision by orders of magnitude are expected.

At low energies, the atom-atom interactions are described by the s-wave scattering length a. The scattering length is the low temperature limit of $-\delta/k$. Therefore, measurements of this type can directly give precise differences of s-wave scattering lengths.[5] Using $\delta=-k\, a$, our current precision would translate to a scattering length difference uncertainty of ±0.7 Å, comparable to the current uncertainty of the caesium triplet scattering length a=1291.2(5) Å.[25] A future accuracy of 100 µrad for Φ yields ±0.009 Å, or 7 ppm. However, the caesium triplet scattering length is so large that k a>1 for even E=1µK × $k_B$. Therefore, our sensitivity to the caesium interatomic potentials at these energies is not so simple and theoretical work is required to establish the sensitivity. Preliminary work has shown a sensitivity to scattering lengths of 1-100 ppm for a measurement accuracy of 100 µrad.[19] Chin and Flambaum[14] have recently suggested that highly sensitive measurements of scattering lengths near Feshbach resonances will set stringent limits on the time variation of the electron-proton mass ratio, a fundamental constant of physics. Near a Feshbach resonance, the phase shift in Fig. 2a has a resonant structure and varies by π. Our technique can accurately measure the phase shift throughout a resonance.

We have demonstrated a fundamentally new scattering method that directly observes the phase shift of an atomic coherence due to quantum scattering. Using atomic clock techniques, we accurately measure the difference of s-wave phase shifts. The technique is quite general; for any atom with a magnetic field insensitive transition, a variety of scattering channels can be explored with high accuracy. For caesium, any of the 16 hyperfine states colliding with the clock coherence as a function of collision energy, magnetic field, and partial angular-momentum wave can be measured. The technique offers direct and unambiguous differences of quantum scattering phase shifts and stringently probes ultra-cold atom-atom interactions.

**Methods**

**Juggling Atomic Fountain.** Here we describe the changes to our juggling fountain that is based on a double magneto-optic trap (MOT).[15,17] We cool and optically pump each cloud of atoms after they are launched from the ultra-high vacuum MOT in a three-dimensional moving-frame optical lattice with degenerate Raman-sideband cooling.[28,29] This optically pumps the atoms into the $|3,3\rangle$ state and cools the atoms to a temperature of typically 500 nK for Cloud 1, and 250 nK for Cloud 2. Instead of launching both clouds with the same velocity, we launch Cloud 1 with a slightly larger velocity than Cloud 2 so that the two clouds collide after their apogees. The two clouds finish passing through one another just before they return downwardly through the microwave cavity. This has two advantages: 1) the two clouds are further separated at launch so that more of the atoms in Cloud 1 survive the launch of Cloud 2, and 2) it shortens the time between the collisions and the detection. After the clouds collide, the atoms spread out spherically with a velocity of $v_r/2 \approx 5$ cm/s. For the scattered atoms to be detected, they must pass through the 1.8 cm cavity apertures and be illuminated by the 2 cm diameter detection laser beam. The highest collision rate occurs when the two cloud centres coincide, and the time from this point until we detect the scattered atoms is typically 0.13 s, so that these atoms spread to a diameter of 1.3 cm.

We have added six 9.2 GHz microwave cavities to our juggling fountain to drive micro-wave transitions. The cavities used for the clock pulses and the state preparation are $TE_{011}$ cylindrical cavities, dielectrically loaded with fused silica to reduce their size. The state-preparation cavities have 12 mm diameter apertures in their endcaps, through which the atoms pass, and those for the clock cavity are 18 mm diameter. Above the clock cavity, there is a $TE_{0,1,13}$ cavity that we use to probe the magnetic field,[11] which is maintained near 15 mG. We pulse the microwaves to the clock cavity and, because of some leakage to the $TE_{0,1,13}$ cavity, we pulse the appropriate power and phase to it. We adjust the amplitude and phase so that the phase of the microwave field in the clock cavity is constant to less than 0.015 rad up to 7.4 mm above the centre of the clock cavity. In addition, we apply the 5ms long pulses when the atoms are centred in the clock cavity. In a future version, we will eliminate this leakage.

For the final state detection after the atoms pass downwardly through the clock cavity, we insert an aperture in the detection laser beam to limit the contributions from the unscattered atoms in Clouds 1 and 2. With a vertical aperture height of order 1 cm, the background Ramsey fringes (violet, red, and green curves in Fig. 2) are small. With no aperture, the background Ramsey fringes in Fig. 2 have amplitudes twice as large as the amplitude for scattered atoms, and still give the same phase shift.

**State Preparation.** Both clouds of atoms are launched from the optical lattice in the $|3,3\rangle$ state and then pass through four microwave state-preparation cavities. A typical sequence for T=233 ms follows whereas, for other launch velocities, the order of the pulses may be slightly different. To minimize backgrounds, we purify the optical pumping by transferring any residual atoms in $|3,1\rangle$ and $|3,2\rangle$ to the F=4 hyperfine level with composite π pulses[30] in the first state-preparation cavity and pulsing a laser beam that pushes F=4 atoms from the fountain. This is repeated in the second state-preparation cavity for $|3,-1\rangle$ and $|3,0\rangle$. After this, Cloud 2 enters the first cavity and, as it travels through the first three cavities, a composite π pulse in each cavity transfers the atoms in $|3,3\rangle$ to $|4,3\rangle$ to $|3,2\rangle$, and finally to $|4,1\rangle$. In the middle of this sequence, Cloud 1 is in the third cavity and we transfer those atoms from $|3,3\rangle$ to $|4,4\rangle$. After both clouds leave the cavities, we apply a laser pulse to repump any atoms left in F=3. A stimulated-Raman transition then velocity selectively transfers the atoms in Cloud 2 from $|4,1\rangle$ to $|3,0\rangle$ before they enter the clock microwave cavity. The peak densities of Cloud 1(2) at launch are $6\times10^9 (1.2\times10^9)$ cm$^{-3}$ which we measure by observing collisions within one cloud[17] between $|3,0\rangle$ and $|4,0\rangle$, for which the triplet scattering length[25] gives a cross-section near the unitary-limit. The number of atoms in each cloud is about $1.6\times10^9$ and $3\times10^8$.

The authors declare no competing financial interests. Correspondence and requests for materials should be addressed to K. G. (kgibble@psu.edu).

[1] Hall, D. S., Matthews, M. R., Ensher, J. R., Wieman, C. E., & Cornell, E. A., Dynamics of Component Separation in a Binary Mixture of Bose-Einstein Condensates, Phys. Rev. Lett. **81**, 1539 - 1542 (1998).

[2] Stenger, J., *et al.*, Spin domains in ground-state Bose–Einstein condensates, Nature **396**, 345-348 (1998).

[3] Khaykovich, L., *et al.*, Formation of a Matter-Wave Bright Soliton, Science **296**, 1290 – 1293 (2002).

[4] Strecker, K. E., Partridge, G. B., Truscott, A. G., & Hulet, R. G., Formation and propagation of matter-wave soliton trains, Nature **417**, 150-153 (2002).

[5] Widera, A. *et al.*, Precision measurement of spin-dependent interaction strengths for spin-1 and spin-2 $^{87}$Rb atoms, New J. Phys. **8,** 152 (2006).

[6] DeMarco, B., & Jin, D. S., Onset of Fermi Degeneracy in a Trapped Atomic Gas, Science **285**, 1703 – 1706 (1999).

[7] Modugno, G., *et al.*, Collapse of a Degenerate Fermi Gas, Science **297**, 2240 – 2243 (2002).

[8] O'Hara, K. M., Hemmer, S. L., Gehm, M. E., Granade, S. R., Thomas, J. E., Observation of a Strongly Interacting Degenerate Fermi Gas of Atoms, Science **298**, 2179 – 2182 (2002).

[9] Gupta, S., *et al.*, Radio-Frequency Spectroscopy of Ultracold Fermions, Science **300**, 1723 – 1726 (2003).

[10] Gibble, K., & Chu, S., Laser-Cooled Cs Frequency Standard and a Measurement of the Frequency Shift due to Ultracold Collisions, Phys. Rev. Lett. **70**, 1771-1774 (1993).

[11] Fertig, C., & Gibble, K., Measurement and Cancellation of the Cold Collision Shift in an $^{87}$Rb Fountain Clock Phys. Rev. Lett. **85**, 1622-1625 (2000).

[12] Wynands, R., & Weyers, S., Atomic fountain clocks, Metrologia **42**, S64-S79 (2005).

[13] Inouye, S., *et al.*, Observation of Feshbach resonances in a Bose–Einstein condensate, Nature **392**,151-154 (1998).

[14] Chin, C., & Flambaum, V.V., Enhanced Sensitivity to Fundamental Constants In Ultracold Atomic and Molecular Systems near Feshbach Resonances, Phys. Rev. Lett. **96**, 230801 (2006).

[15] Legere, R., & Gibble, K., Quantum Scattering in a Juggling Atomic Fountain, Phys. Rev. Lett. **81**, 5780-5783 (1998).

[16] Kasevich, M. *et al.*, Atomic velocity selection using stimulated Raman transitions, Phys. Rev. Lett. **66**, 2297 - 2300 (1991).

[17] Gibble, K., Chang, S., & Legere, R., Direct Observation of s-wave Atomic Collisions, Phys. Rev. Lett **75**, 2666-2669 (1995).

[18] Legere, R. J., Ph.D. thesis, Quantum Scattering in a Juggling Atomic Fountain, Yale University, unpublished (1999).

[19] S. J. J. M. F. Kokkelmans, Ph.D. thesis, Interacting Atoms in Clocks and Condensates, University of Eindhoven, unpublished (2000).

[20] Monroe, C. R., Cornell, E. A., Sackett, C. A., Myatt, C. J., & Wieman, C. E., Measurement of Cs-Cs elastic scattering at T=30 μK, Phys. Rev. Lett. **70**, 414-417 (1993).

[21] Myatt, C. J., Burt, E. A., Ghrist, R. W., Cornell, E. A., & Wieman, C. E., Production of Two Overlapping Bose-Einstein Condensates by Sympathetic Cooling, Phys. Rev. Lett. **78**, 586 - 589 (1997).

[22] Ensher, J. R., Jin, D. S., Matthews, M. R., Wieman, C. E., and Cornell, E. A., "Bose-Einstein Condensation in a Dilute Gas: Measurement of Energy and Ground-State Occupation," Phys. Rev. Lett. **77**, 4984-4987 (1996).

[23] DePue, M. T., McCormick, C., Winoto, S. L., Oliver, S., and Weiss, D.S., Unity Occupation of Sites in a 3D Optical Lattice, Phys. Rev. Lett. **82**, 2262 - 2265 (1999).

[24] Chin, C., Vuletic, V., Kerman, A. J., & Chu, S., High Resolution Feshbach Spectroscopy of Cesium, Phys. Rev. Lett. **85**, 2717-2720 (2000).

[25] Chin, C., *et al.*, Precision Feshbach spectroscopy of ultracold Cs$_2$," Phys. Rev. A **70**, 032701 (2004).

[26] Thomas, N. R., Kjærgaard, N., Julienne, P. S., & Wilson, A. C., Imaging of s and d Partial-Wave Interference in Quantum Scattering of Identical Bosonic Atoms., Phys. Rev. Lett. **93**, 173201 (2004).

[27] Buggle, Ch., Léonard, J., von Klitzing, W., & Walraven, J. T. M., Interferometric Determination of the s and d-Wave Scattering Amplitudes in $^{87}$Rb., Phys. Rev. Lett. **93**, 173202 (2004).

[28] Hamann, S. E., *et al.,* Resolved-Sideband Raman Cooling to the Ground State of an Optical Lattice, Phys. Rev. Lett. **80**, 4149-4152 (1998).

[29] Treutlein, P., Chung, K. Y., & Chu, S., High-brightness atom source for atomic fountains, Phys. Rev. A **63**, 051401(R) (2001).

[30] Levitt, M. H., Composite pulses, Prog. in Nuclear Magnetic Resonance Spectroscopy **18**, 61-122 (1986).

The authors gratefully acknowledge stimulating discussions with Boudewÿn Verhaar and Servaas Kokkelmans and contributions from Ruoxin Li. This work was supported by NASA, NSF, ONR, and Penn State University.